\documentclass{iaus}
\usepackage{graphicx}

\title[Aligning VLBI and GAIA celestial reference frames]{VLBI observations of weak extragalactic radio sources for the alignment of the future GAIA frame with the ICRF}

\author[G. Bourda et al.]{G. Bourda$^1$,P. Charlot$^1$, R. Porcas$^2$, and S. Garrington$^3$}

\affiliation{$^1$Laboratoire d'Astrophysique de Bordeaux, Universit\'{e} Bordeaux 1, CNRS, Floirac, France 
 \break{email: bourda@obs.u-bordeaux1.fr}                                \\[\affilskip]
$^2$Max Planck Institute for Radio Astronomy, Bonn, Germany  \\[\affilskip]
$^3$Jodrell Bank Observatory, The University of Manchester, Macclesfield, UK}

\pubyear{2008}
\volume{248}  
\pagerange{1--2}
\date{?? and in revised form ??}
\jname{A Giant Step: from Milli- to Micro-arcsecond Astrometry}
\editors{W.-J. Jin, I. Platais, M. Perryman, eds.}

\begin{document}

\maketitle

\begin{abstract}
The space astrometry mission GAIA will construct a dense optical QSO-based celestial reference frame. For consistency between the optical and radio positions, it will be important to align the GAIA frame and the International Celestial Reference Frame (ICRF) with the highest accuracy. Currently, it is found that only 10\% of the ICRF sources are suitable to establish this link, either because they are not bright enough at optical wavelengths or because they have significant extended radio emission which precludes reaching the highest astrometric accuracy. In order to improve the situation, we have initiated a VLBI survey dedicated to finding additional high-quality radio sources for aligning the two frames. The sample consists of about 450 sources, typically 20 times weaker than the current ICRF sources, which have been selected by cross-correlating optical and radio catalogues. This paper presents the observing strategy and includes preliminary results of observation of 224 of these sources with the European VLBI Network in June 2007.
\end{abstract}


The ICRF (International Celestial Reference Frame) is the fundamental celestial reference frame adopted by the International Astronomical Union (IAU) in August 1997 (\cite{Ma98}; \cite{Fey04}), currently based on the VLBI (Very Long Baseline Interferometry) positions of 717 extragalactic radio sources. 
The European space astrometry mission GAIA, to be launched by 2011, will survey about one billion stars in our Galaxy and 500~000 Quasi Stellar Objects (QSOs), brighter than magnitude 20 (\cite{Perryman01}). Unlike Hipparcos, GAIA will construct a dense optical celestial reference frame directly in the visible, based on the QSOs with the most accurate positions (i.e. with magnitude $V\leq18$; \cite{Mignard03}). 
In the future, the alignment of the ICRF and the GAIA frame will be crucial for ensuring consistency between the measured radio and optical positions. This alignment, to be determined with the highest accuracy, requires several hundreds of common sources, with a uniform sky coverage and very accurate radio and optical positions. Obtaining such accurate positions implies that the link sources must have $V\leq18$ and no extended VLBI structures. 
In a previous study, we investigated the current status of this link based on the present list of ICRF sources (\cite{Bourda07}). We found that although about 30\% of the ICRF sources have an optical counterpart with $V \leq 18$, only one third of these are compact enough on VLBI scales for the highest astrometric accuracy. 
Overall, only 10\% of the current ICRF sources ($\simeq$70~sources) are thus available today for the alignment of the GAIA frame. This highlights the need to identify additional suitable radio sources, which is the purpose of the project described here.

Searching for additional sources suitable for aligning accurately the ICRF and the GAIA frame implies going to weaker radio sources having flux densities typically below 100~mJy. This can now be realized owing to recent increases in the VLBI network sensitivity (e.g. recording now possible at 1Gb/s) and by using a network with large antennas like the EVN (European VLBI Network).
A sample of about 450 radio sources, for which there are no published VLBI observations, was selected for this purpose by cross-identifying the NRAO VLA Sky Survey (NVSS; \cite{Condon98}), a deep radio survey (complete to the 2.5~mJy level) that covers the entire sky north of $-40^{\circ}$, with the \cite{Veron06} optical catalogue. This sample is based on the following criteria: (i) $V\leq18$ (to ensure very accurate positions with GAIA), (ii) $\delta\geq-10^{\circ}$ (for possible observing with northern VLBI arrays), and (iii) NVSS flux density~$\geq$~20~mJy (for possible VLBI detection). 
The observing strategy to identify appropriate link sources in the sample includes three successive steps: (1) To determine the VLBI detectability of these weak radio sources; (2) To image and determine an accurate astrometric position for the sources detected in the previous step; and (3) To refine the astrometry for the most compact sources of the sample. 

\begin{figure}
\centering
  \includegraphics[scale=0.18,angle=-90]{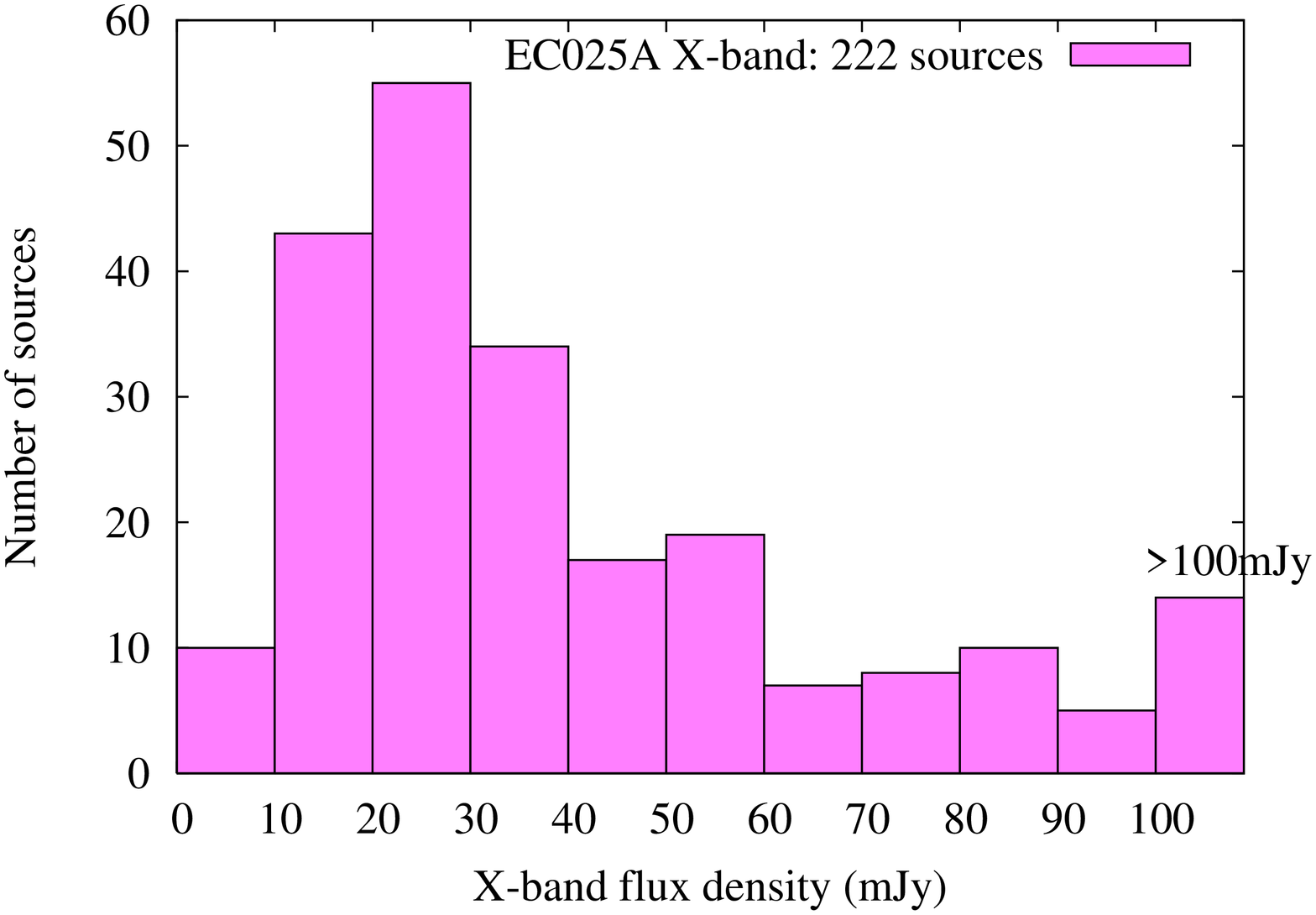}
  \includegraphics[scale=0.18,angle=-90]{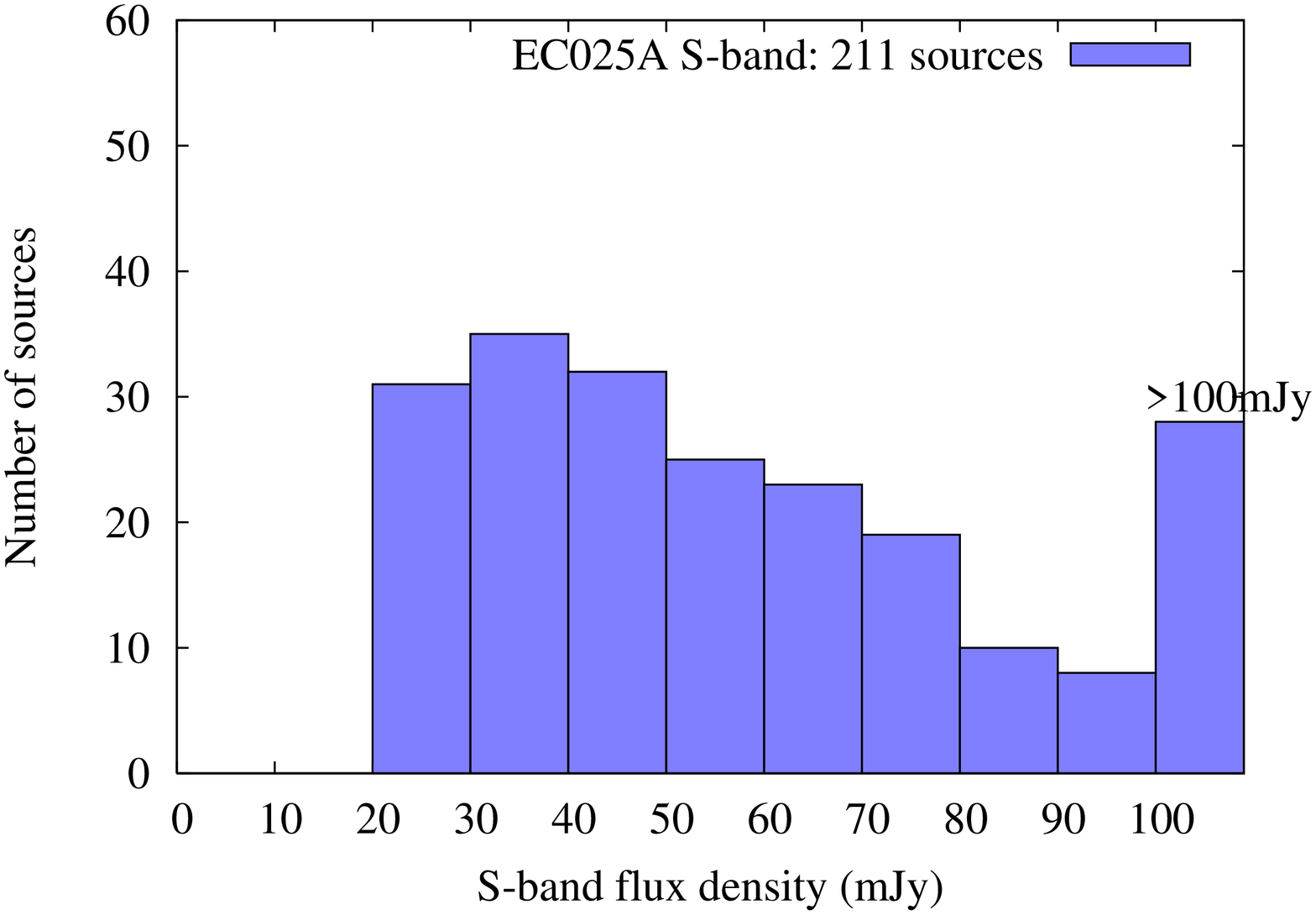}
\caption{Flux density distribution (units in mJy) for the sources detected in our initial experiment EC025A (June 2007). On average, the flux density of these sources is 20~times and 7~times weaker than that of the ICRF and VCS sources (\cite{Kovalev07} and references therein).}\label{fig:Fig1}
\end{figure}

The first observations for this project (experiment EC025A) were made in June 2007, with a network of four EVN telescopes. Half of our sample (i.e. 224~target sources, most of which belonging to the CLASS catalogue; \cite{Myers03}) was observed during this experiment to determine their VLBI detectability (step 1 described above). The rest of the sources will be observed in October 2007. 
Our results for EC025A indicate excellent detection rates of 99\% at X band and 95\% at S band, with 222~sources and 211~sources detected at X and S bands, respectively. The mean correlated flux densities have a median value of 32~mJy at X band and 55~mJy at S band (see Figure~\ref{fig:Fig1}). The spectral index $\alpha$ of the 211 radio sources detected at both frequencies was also investigated. Its median value is $-0.3$ and most of the sources have $\alpha > -0.5$, hence indicating that they must have a dominating core component, which is very promising for the future stages of this project.

\vspace{-0.3cm}

\end{document}